\documentclass[aps,prd,floats,floatfix,twocolumn,nofootinbib,amssymb,amsmath]
{revtex4}

\usepackage{graphicx}
\usepackage{bm}

\newcommand{\beq}{\begin{equation}}
\newcommand{\eeq}{\end{equation}}
\newcommand{\beqa}{\begin{eqnarray}}
\newcommand{\eeqa}{\end{eqnarray}}

\begin{document}

\title{Model Waveform Accuracy Standards for Gravitational Wave Data Analysis}

\author{Lee Lindblom${}^1$, Benjamin J. Owen${}^2$, and Duncan A. Brown${}^3$}

\affiliation{${}^1$ Theoretical Astrophysics 130-33, California Institute of
Technology, Pasadena, CA 91125}

\affiliation{${}^2$ Institute for
Gravitation and the Cosmos,
and Center for Gravitational Wave Physics, 
Department of Physics, The Pennsylvania State University,
University Park, PA 16802}

\affiliation{${}^3$ Department of Physics, Syracuse University,
Syracuse, NY 13244}

\begin{abstract}
Model waveforms are used in gravitational wave data analysis to detect
and then to measure the properties of a source by matching the model
waveforms to the signal from a detector.  This paper derives accuracy
standards for model waveforms which are sufficient to ensure that
these data analysis applications are capable of extracting the full
scientific content of the data, but without demanding excessive
accuracy that would place undue burdens on the model waveform
simulation community. These accuracy standards are intended
primarily for broad-band 
model waveforms produced by numerical simulations, but
the standards are quite general and apply equally to such waveforms
produced by analytical or hybrid analytical-numerical methods.
\end{abstract}
\date{\today}

\maketitle

\section{Introduction}
\label{s:Introduction}

The purpose of this paper is to derive standards for the accuracy of
model waveforms sufficient to ensure that those waveforms are good
enough for their intended uses in gravitational wave data analysis.
This is a timely and important subject which has received relatively
little attention in the scientific literature up to this point.
Several gravitational wave detectors~\cite{Abbott:2007kva,
Acernese:2008zz, Grote:2008zz}  have now achieved a high enough level
of sensitivity that the first astrophysical observations are expected
to occur within the next few years.   The numerical relativity
community has also matured to the point that several groups are now
computing model gravitational waveforms for the inspiral and merger of
black hole and neutron star binary systems~\cite{Shibata2002,
Pretorius2005a, Campanelli2006a, Faber2006, Baker2006a, Diener2006,
Loffler2006, Herrmann2007b, Bruegmann2006,
Scheel2008}. Beyond the pioneering work of Mark
Miller~\cite{Miller2005} and Stephen Fairhurst~\cite{Fairhurst2008}, 
however, little effort has gone into
thinking about the question of how accurate these model waveforms need
to be.

This paper contributes to this discussion by formulating
a set of accuracy standards for model waveforms, sufficient to ensure
that those waveforms are able to fulfill the detection and
parameter-measurement roles they will be required to play in
gravitational wave data analysis.  The standards presented here are
designed to be optimal in the sense that waveforms of lesser accuracy
would result in some loss of scientific information from the data,
while more accurate waveforms would merely increase the cost of
computing the waveforms without increasing their scientific value in
data analysis.  Our discussion is done here at a fairly abstract
level, with the intention that these accuracy standards should be
applicable to model waveforms produced by approximate analytical
methods (such as post-Newtonian 
expansions) as well as model waveforms produced by
numerical simulations.

Our discussion is divided into two parts: The first part, in
Sec.~\ref{s:IdealDetectorCase}, assumes that the calibration of the
gravitational wave detector is perfect.  That is, we assume that the
response function used to convert the interferometer output to a
gravitational wave signal is known exactly.  In this ideal detector
case, we present simple derivations for the needed accuracy of model
waveforms for detection and separately for parameter measurement
purposes.  The second part, in
Sec.~\ref{s:IncludingCalibrationErrors}, evaluates the effects of
calibration error (i.e., the errors in the measurement of the response
function) on the needed accuracy requirements for model waveforms.

\begin{figure}
\centerline{\includegraphics[width=3in]{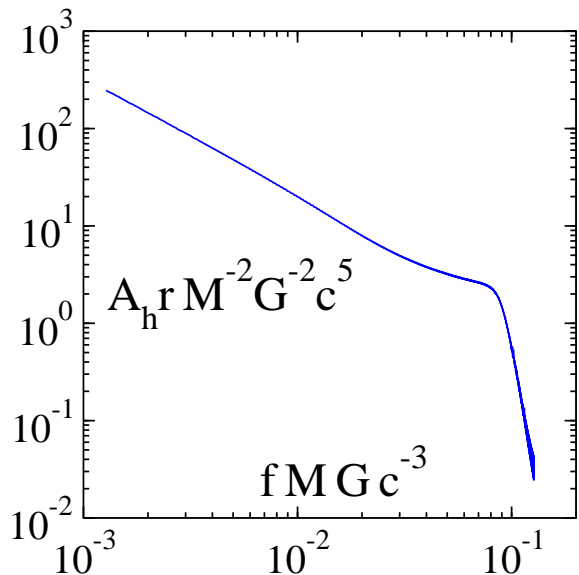}}
\caption{\label{f:FFThAmp} Amplitude $A_h$ of the frequency-domain
  gravitational waveform for an equal-mass non-spinning binary
  black-hole system.  Constants representing the total mass, $M$,
  and the distance, $r$, to the binary system have been introduced to
  make the graphed quantities dimensionless, independent
  of the mass of the source, and (asymptotically) independent 
  of the observer's position; $G$ is Newton's
  constant and $c$ is the speed of light.
}
\end{figure}

\begin{figure}
\centerline{\includegraphics[width=3in]{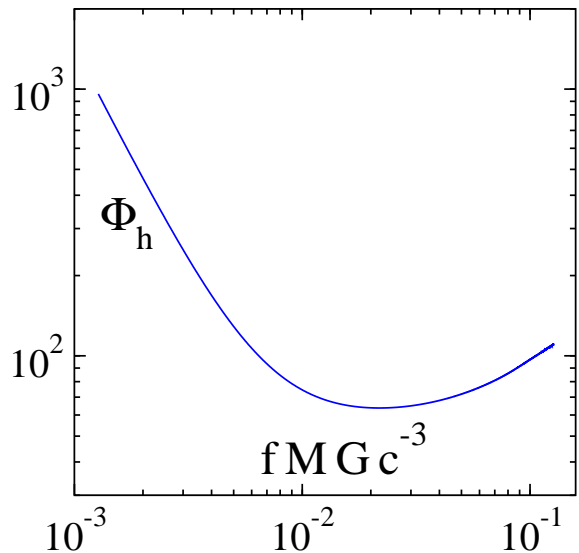}}
\caption{\label{f:FFThPhase} Phase $\Phi_h$ (measured in
radians) of the frequency-domain gravitational waveform for an
equal-mass non-spinning binary black-hole system.  The constant
$M$, representing the total mass of the binary system, has been
introduced to express the frequency in dimensionless
units, and to make the graphed quanties independent of the
mass of the source; $G$ is Newton's constant and $c$ is the speed of light.}
\end{figure}

For simplicity, all of our discussion here will be based on the
frequency-domain representations of gravitational waveforms $h(f)$,
defined as,
\begin{eqnarray}
h(f)=\int_{-\infty}^{\infty} h(t)e^{-2\pi i f t} dt,
\end{eqnarray}
where $h(t)$ is the time-domain representation of the
waveform.\footnote{We follow the convention of the LIGO Scientific
Collaboration~\cite{T010095} (and the signal-processing community)
by using the phase factor $e^{-2\pi i f
t}$ in these Fourier transforms; most of the early gravitational
wave literature and essentially all other computational physics
literature use $e^{2\pi i f t}$.  This choice does not affect any of
the subsequent equations in this paper.}  Since the frequency-domain
representation of waveforms is somewhat less familiar to the numerical
relativity community, we include Figs.~\ref{f:FFThAmp} and
\ref{f:FFThPhase} to illustrate the frequency-domain amplitude $A_h$
and phase $\Phi_h$, defined as $h(f)=A_he^{i\Phi_h}$, of the waveform
for a binary black-hole system composed of equal-mass non-spinning
holes.  The numerical part of this waveform was produced by the
Caltech/Cornell numerical relativity group~\cite{Scheel2008},
and a post-Newtonian model waveform was stitched on for the lowest part of the
frequency range~\cite{Boyle2008b}.   The constants $M$ and $r$
(used as scale factors in these figures) are, respectively, the total
mass and the luminosity distance of the binary system. 
For clarity of presentation, we have removed the linear in $f$ part of
$\Phi_h(f)$, which corresponds to shifting the origin of the time
coordinate.  We have also adjusted the constant part of the phase so
that $\Phi_h(f)$ does not have a zero and can be graphed more
conveniently.  These time and phase constants are kinematic
parameters, which are not related to the internal dynamics of the
waveforms.
The resulting $\Phi_h(f)$ shown in Fig.~\ref{f:FFThPhase} is not
monotonic in $f$, and this might seem counterintuitive at first.  We
note that the addition of the linear in $f$ term, $2\pi t_0 f$, can
make $\Phi_h(f)$ either monotonic increasing or decreasing, depending
on the value of the time constant $t_0$.  We also note that the $\Phi_h(f)$
illustrated in Fig.~\ref{f:FFThPhase} is the phase of the Fourier
transform of $h(t)$, and consequently has no simple relationship
with the phase of the time-domain waveform.

While the accuracy standards derived here apply to any type of
gravitational waveform, in practice we expect them to be most useful
for broad-band waveforms such as those from compact binary systems.
Narrow band gravitational wave signals, such as from rotating neutron
stars, are expected to be analyzed using modified versions of the data
analysis methods described here, i.e., using the
$\mathcal{F}$-statistic matched filter or time-domain
heterodyning~\cite{Abbott:2003yq}.  The accuracy-standard analysis
presented here could easily be extended to these
cases~\cite{Jaranowski:1998ge, Dupuis:2005xv}.  However narrow-band
continuous-wave sources have very simple and robust phenomenological
waveforms, and so the question of modeling error does not arise in the
same way it does for systems with very complex waveforms like compact
binary systems.

\section{Ideal Detector Case}
\label{s:IdealDetectorCase}

We split the discussion of the ideal detector case into three
parts.  First we present in Sec.~\ref{s:ParameterMeasurementIdeal} a
simple derivation of the accuracy of model waveforms needed to ensure
no loss of scientific information when the waveform is used to measure
the physical properties of a gravitational wave signal.  Second we
present in Sec.~\ref{s:SignalDetectionIdeal} a simple condition on the
accuracy of model waveforms needed to ensure a prescribed level of
detection efficiency.  These accuracy requirements in
Secs.~\ref{s:ParameterMeasurementIdeal} and
\ref{s:SignalDetectionIdeal} are optimal in the sense that any model
waveform violating them would decrease the scientific effectiveness of
the detector.  Unfortunately these conditions depend on the detector's
noise curve in a complicated way, so enforcing them is somewhat
complicated.  Therefore we present in
Sec.~\ref{s:SufficientConditionsIdeal} a set of simpler conditions
that are nevertheless sufficient to guarantee that the optimal conditions
are satisfied.  While these sufficient conditions are somewhat
stronger than needed, they are much simpler to apply; so we hope they
will be easier for the waveform simulation community to adopt and use on
a regular basis. 

\subsection{Accuracy Requirements for Measurement}
\label{s:ParameterMeasurementIdeal}

The question we wish to address here is, how small must the difference
between two waveforms be to ensure that measurements
with a particular detector are unable to distinguish them?  This
condition determines how accurate a model waveform $h_m$ must be to make
it indistinguishable from the exact physical waveform $h_e$ through any
measurement with a particular detector.

For simplicity, we will perform our analysis in terms of the
frequency-domain representation of the waveforms $h_e(f)$ and
$h_m(f)$.  Consider the one parameter family of waveforms,
\begin{eqnarray}
h(\lambda,f) &=& (1-\lambda) h_e(f) + \lambda h_m(f),\nonumber\\
&\equiv& h_e(f) + \lambda \delta h(f),
\label{e:FamilyDef}
\end{eqnarray}
that interpolates between $h_e$ and $h_m$ as $\lambda$ varies between
$0$ and $1$.  We now ask the related question, how accurately can a
particular gravitational wave detector measure this parameter
$\lambda$?  There exists a well developed theory of parameter
measurement accuracy for gravitational wave data analysis, discussed
for example in Finn~\cite{Finn1992}, Finn and Chernoff~\cite{Finn1993}
and Cutler and Flanagan~\cite{CutlerFlanagan1994}.  We construct the
noise-weighted inner product $\langle h_e | h_m \rangle$ given by
\begin{eqnarray}
\langle h_e | h_m \rangle = 2
\int_{0}^\infty \frac {h_e^*(f) h_m(f) + h_e(f) h_m^*(f)}{S_n(f)}df,
\label{e:innerproduct}
\end{eqnarray}
where $S_n(f)$ is the one-sided power spectral density of the
detector strain noise.  
In defining this inner product, we use the fact that the
Fourier transform of the \emph{real} gravitational-wave strain $h(t)$
satisfies $h(f) = h^*(-f)$, thus allowing us to define the inner
product as an integral over positive frequencies only.

The variance $\sigma_\lambda^2$ of measurements of the parameter
$\lambda$ is given by the expression
\begin{eqnarray}
\sigma_\lambda^{-2} &=& \left\langle \frac{\partial h}{\partial \lambda}\left|
\frac{\partial h}{\partial \lambda}\right.\right\rangle
= \langle \delta h|\delta h \rangle,
\end{eqnarray}
using Eq.~(3.20) of Ref.~\cite{Finn1992}, or Eq.~(2.8) of
Ref.~\cite{CutlerFlanagan1994}.  If the standard
deviation $\sigma_\lambda$ were
greater than one (the parametric distance between $h_e$ and $h_m$),
then the two waveforms would be indistinguishable through any
measurement with the given detector.  Thus the condition,
$\sigma_\lambda>1$, or equivalently,
\begin{eqnarray}
\langle \delta h|\delta h\rangle < 1,
\label{e:MeasurementLimit1}
\end{eqnarray}
ensures that the two waveforms are indistinguishable.\footnote{This
inequality, obtained through a different argument, was
presented by Stephen Fairhurst in a talk at the ``Interplay between
Numerical Relativity and Data Analysis'' Miniworkshop, at the KITP,
UCSB in January 2008.} If we consider $h_e$ to be the exact waveform, and
$\delta h$ to be the difference between the model and the exact
waveforms, then the model waveform will be indistinguishable from the
exact if and only if $\langle \delta h|\delta h\rangle < 1$.  
Our derivation of this condition is based on the simple one-parameter
family of waveforms defined in Eq.~(\ref{e:FamilyDef}), however, this
argument applies to every possible model waveform error $\delta h$.
So the argument and resulting condition are completely general:
any acceptable waveform model must lie within a ball of unit radius
centered on the exact waveform.

We can re-express this limit, Eq.~(\ref{e:MeasurementLimit1}), as a
simple condition on the needed phase and amplitude accuracies of the
model waveforms.  Let $\chi_e$ and $\Phi_e$ denote real functions
representing the logarithmic amplitude and phase of the exact
waveform: $h_e\equiv e^{\chi_e+i\Phi_e}$.  The model waveform may
differ from the exact in both amplitude and phase: $h_e+\delta h =
h_ee^{\delta \chi+ i\delta \Phi}$ or to first order $\delta h =
(\delta\chi+i \delta \Phi )h_e$.  It is also useful to introduce the
normalized waveform $\hat h_e = h_e \rho^{-1}$, which satisfies
$\langle\hat h_e|\hat h_e\rangle=1$, where $\rho^2 = \langle h_e |
h_e\rangle$.  Using these quantities we can express the inner product
$\langle\delta h|\delta h\rangle$ in the following way
\begin{eqnarray}
\langle\delta h|\delta h\rangle &=&
\rho^2\left\langle\left(\delta\chi + i \delta \Phi\right) \hat h_e\left|
\left(\delta\chi+i\delta \Phi \right)\hat h_e \right.\right\rangle,\nonumber\\
&=& \rho^2\left(\overline{\delta\chi}^2 +\overline{ \delta\Phi}^2\right), 
\label{e:MeasurementLimit2} 
\end{eqnarray}
where the signal-weighted averages of the logarithmic amplitude and 
phase errors are defined by,
\begin{eqnarray}
\overline{\delta\chi}^2&\equiv &
\bigl\langle \delta\chi \hat h_e \bigl|
\delta \chi \hat h_e \bigr\rangle ,
\label{e:AmplitudeAverageDef}\\
\overline{\delta  \Phi}^2 &\equiv&
\bigl\langle \delta\Phi \hat h_e \bigl|\delta \Phi \hat h_e \bigr\rangle.
\label{e:PhaseAverageDef}
\end{eqnarray} 
We can use these definitions to express
Eq.~(\ref{e:MeasurementLimit1}) as a simple limit on the
signal-weighted averages of the logarithmic amplitude and phase
errors:
\begin{eqnarray}
\overline{\delta\chi}^2 +\overline{ \delta\Phi}^2 < \frac{1}{\rho^2}.
\label{e:MeasurementLimit3} 
\end{eqnarray}
Equation~(\ref{e:MeasurementLimit1}) or equivalently
Eq.~(\ref{e:MeasurementLimit3}) are the basic requirements on model
waveforms for measurement purposes.  These requirements are optimal in
the sense that waveforms more accurate than this would not improve
scientific measurements, while less accurate waveforms would degrade
some measurements.

\subsection{Accuracy Requirements for Detection}
\label{s:SignalDetectionIdeal}

The signal-to-noise ratio for the detection of a signal $h_e$ using an
optimal filter constructed from the model waveform $h_m$ is given by
\begin{eqnarray}
\rho_m &=& \langle h_e|\hat h_m\rangle
       = \frac{\langle h_e|h_m\rangle}{\langle h_m|h_m\rangle^{1/2}},
\\\nonumber 
\end{eqnarray}
cf. Eq.~(A24) of Ref.~\cite{CutlerFlanagan1994}.  The question we wish
to address here is, how accurate must the model waveform $h_m$ be to
ensure no significant loss in the efficiency of detecting the signal
$h_e$?  Detections are made when a signal is observed that exceeds a
predetermined threshold signal-to-noise ratio.  So errors in
evaluating the signal-to-noise ratio will decrease the detection
efficiency.  We must determine therefore how errors in the model
waveform $h_m$ degrade the measured signal-to-noise ratio, $\rho_m$,
relative to the optimal signal-to-noise ratio
$\rho=\langle h_e|h_e\rangle^{1/2}$.  We
introduce a parameter $\epsilon$, referred to as the \emph{mismatch}
in the gravitational-wave data-analysis
literature~\cite{Owen:1995tm}, that measures this
signal-to-noise reduction:
\begin{eqnarray}
\rho_m = (1-\epsilon)\rho.
\label{e:SignalToNoiseLoss}
\end{eqnarray}

To determine how model waveform errors $\delta h$ affect $\epsilon$,
we write the model waveform as $h_m=h_e+\delta h$, and re-write
Eq.~(\ref{e:SignalToNoiseLoss}) in terms of the definitions of $\rho$
and $\rho_m$:
\begin{eqnarray}
\label{e:powermismatch}
\frac{\langle h_e|h_e+\delta h\rangle^2}
{\langle h_e+\delta h|h_e+\delta h\rangle} 
= (1-\epsilon)^2\langle h_e| h_e\rangle.
\end{eqnarray}
This equation can be simplified by decomposing $\delta h$ into two
parts: $\delta h_\parallel=\hat h_e\langle\delta h|\hat h_e\rangle$
and $\delta h_\perp=\delta h-\delta h_\parallel$.  This $\delta
h_\parallel$ is proportional to $h_e$, while $\delta h_\perp$ is
orthogonal to it in the sense that $\langle \delta h_\perp|
h_e\rangle=0$.  
Using these expressions it is straightforward to
derive the following relationship between the mismatch $\epsilon$ and
the model waveform error:
\begin{eqnarray}
\epsilon = \frac{\langle \delta h_\perp | \delta h_\perp\rangle}
{2\langle h_e| h_e\rangle}.
\label{e:SignalToNoiseLoss2}
\end{eqnarray}
We have kept only the lowest-order terms in $\delta h$ (which we
assume to be small) in this expression. 
Equation~(\ref{e:SignalToNoiseLoss2}) shows that
the mismatch $\epsilon$ is proportional to the square of the distance
between two waveforms, as measured by the noise-weighted inner product.

To ensure a high level of detection efficiency while using an optimal
filter based on $h_m$, we must ensure that the model waveform error
$\delta h$ is small enough to prevent $\epsilon$ from becoming
unacceptably large.  Let $\epsilon_{\max}$ be the maximum mismatch
compatible with our target detection efficiency.  In that case
Eq.~(\ref{e:SignalToNoiseLoss2}) places the following limit on the
model waveform error:
\begin{eqnarray}
\langle \delta h_\perp | \delta h_\perp\rangle < 2\rho^2 \epsilon_{\max}.
\label{e:DetectionLimit}
\end{eqnarray}
When real searches are conducted to detect signals by matching
to a model waveform, the measured signal-to-noise ratio $\rho_m$ is
maximized over different time and phase offsets of the model waveform.  
Thus the part of the model waveform phase error
linear in $f$ (the part that
depends on the time and phase offsets)
is not relevant for detection.  Strictly speaking then,
the inner product that appears in Eq.~(\ref{e:DetectionLimit})
should be interpreted 
as the ``match'' inner product, defined as
the Eq.~(\ref{e:innerproduct}) inner product
optimized over these time and phase
offsets~\cite{Owen:1995tm}.

Equation~(\ref{e:DetectionLimit}), with 
$\langle \delta h_\perp|\delta h_\perp\rangle$
interpreted as the match inner product,
gives the optimal condition on the
allowed model waveform error for detection purposes.  Unfortunately it
is not generally possible to determine what the orthogonal part of the
waveform error $\delta h_\perp$ actually is without knowing the exact
waveform, so a simpler, easier to evaluate limit is desirable.  We can
obtain such a condition by noting that $\langle \delta h|\delta
h\rangle \geq \langle \delta h_\perp |\delta h_\perp\rangle$
(where the inner product on the left can be the standard
noise-weighted inner product); so a
sufficient condition that ensures the target detection efficiency is
\begin{eqnarray}
\langle \delta h | \delta h\rangle < 2\rho^2 \epsilon_{\max}.
\label{e:DetectionLimit1}
\end{eqnarray}
We note that this condition is considerably weaker (depending on the
values of $\rho$ and $\epsilon_{\max}$) than the limit presented in
Eq.~(\ref{e:MeasurementLimit1}) to ensure no loss of accuracy in
parameter measurements. We can also transform this limit,
Eq.~(\ref{e:DetectionLimit1}), into a simple expression for the needed
accuracy of the amplitude and phase of model waveforms, in analogy
with those found for measurement accuracy in
Sec.~\ref{s:ParameterMeasurementIdeal}.  As before we express the
model waveform error in terms of logarithmic amplitude and phase
errors: $\delta h =(\delta \chi +i\delta \Phi )h_e$.  Substituting
this into Eq.~(\ref{e:DetectionLimit1}), we arrive at a simple
expression for the limit on the signal-weighted averages of the
logarithmic amplitude and phase errors required for detection:
\begin{eqnarray}
\overline{\delta \chi}^2 +
\overline{\delta \Phi}^2 < 2\epsilon_{\max}.
\label{e:DetectionAmpPhaseLimit}
\end{eqnarray}

The value of the maximum mismatch $\epsilon_{\max}$ that appears in
Eqs.~(\ref{e:DetectionLimit1}) and (\ref{e:DetectionAmpPhaseLimit})
must be set by the demands of the particular data-analysis
application.  Setting $\epsilon_{\max}=0.035$ in a search using a
single model-waveform template, for example, would result in a
reduction in detection rate (for sources that are uniformly distributed
in space) of $1-(1-\epsilon_{\max})^3\approx 0.10$, a target that is
often adopted in LIGO searches for compact binary
inspirals~\cite{Abbott:2003pj, Abbott:2005pe, Abbott:2005qm,
Abbott:2007xi}.  

\begin{figure}
\centerline{\includegraphics[width=3in]{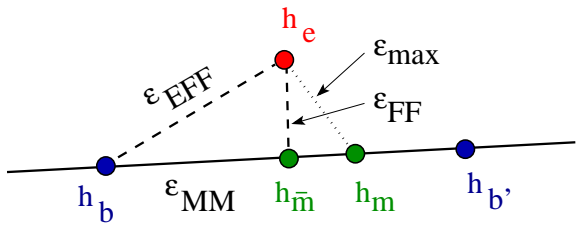}}
\caption{\label{f:Mismatch} Solid line illustrates the model-waveform
submanifold, with particular members of a discrete template bank
$h_{b}$ and $h_{b'}$.  The model waveform $h_{\bar m}$ has the maximum
mismatch $\epsilon_\mathrm{MM}$ with the template bank waveform
$h_{b}$.  The closest exact waveform $h_e$ has mismatch
$\epsilon_\mathrm{FF}$ with $h_{\bar m}$ and total effective mismatch
$\epsilon_\mathrm{EFF}=\epsilon_\mathrm{MM}+\epsilon_\mathrm{FF}$ with
$h_{b}$. The model waveform $h_m$ having the same physical
parameters as $h_e$ may differ from $h_{\bar m}$ due to
modeling errors, and its mismatch
$\epsilon_{\max}$ will always exceed $\epsilon_\mathrm{FF}$:
$\epsilon_{\max}\geq\epsilon_\mathrm{FF}$.}
\end{figure}

Real gravitational-wave searches are more complicated, and making the
appropriate choice of $\epsilon_{\max}$ is more subtle.  Real searches
are generally performed by matching discrete template banks of model
waveforms with the data. Let $\hat h_b$ represent one of the
(normalized) waveforms in the discrete template bank; let 
$\hat h_{\bar m}$
be a (normalized) model waveform whose (appropriately defined)
distance from $\hat h_b$ is the maximum for
the given template bank; and let $\hat h_e$
denote the (normalized) exact waveform whose distance is closest 
to $\hat h_{\bar m}$, see Fig.~\ref{f:Mismatch}.  
The model waveform $\hat h_{m}$ having the same physical
parameters as $\hat h_e$ may not be identical to $\hat h_{\bar m}$, since
there may be a component of the waveform error tangent to the
model-waveform submanifold.
The quantity
$FF=\langle \hat{h}_e | \hat{h}_{\bar m} \rangle\equiv
1-\epsilon_\mathrm{FF}$ is often referred to as the fitting
factor~\cite{Apostolatos1994}; $MM=\langle \hat{h}_{\bar m} | \hat{h}_b
\rangle=1-\epsilon_\mathrm{MM}$ is the minimal
match~\cite{Owen:1995tm}; and we refer to $EFF=\langle \hat{h}_e |
\hat{h}_b \rangle=1-\epsilon_\mathrm{EFF}$ as the ``effective fitting
factor.''  The goal is to have any physical waveform match some
model waveform in the template bank with a mismatch that is no greater
than the chosen $\epsilon_\mathrm{EFF}$ (for example
$\epsilon_\mathrm{EFF}=0.035$).    The value of
$\epsilon_\mathrm{FF}$ is completely determined by the chosen target
$\epsilon_\mathrm{EFF}$, and the parameter $\epsilon_\mathrm{MM}$
which describes the spacing of models in the discrete template bank.
Since $\hat h_{\bar m}$ 
is the best-fit model waveform for the exact waveform
$\hat h_e$, it follows that the relative waveform vector $\hat h_e -
\hat h_{\bar m}$ will be orthogonal to any vector tangent to the
model-waveform submanifold at $\hat h_{\bar m}$; 
thus $0=\langle \hat h_e -
\hat h_{\bar m} | \hat h_b - \hat h_{\bar m}\rangle$ in the limit that 
$\hat h_b$
and $\hat h_e$ are infinitesimally close to $\hat h_{\bar m}$.  
Writing out
this orthogonality condition in terms of the mismatch parameters
defined above gives
\begin{eqnarray}
\epsilon_\mathrm{FF}=\epsilon_\mathrm{EFF}-\epsilon_\mathrm{MM},
\end{eqnarray}  
a kind of Pythagorean theorem for model-waveform mismatches.
(Recall from Eq.~[\ref{e:SignalToNoiseLoss2}] 
that the mismatch $\epsilon$ measures the square of the
distance between nearby waveforms.)  The template banks used for
compact binary searches in initial LIGO are constructed with
$\epsilon_\mathrm{MM}=0.03$
\cite{Abbott:2003pj,Abbott:2005pe,Abbott:2007xi}, implying that the
needed accuracy of the model waveforms is
$\epsilon_\mathrm{FF}=0.005$ when $\epsilon_\mathrm{EFF}=0.035$.
Since $\hat h_{\bar m}$ is the closest model waveform to $\hat
h_e$, it follows that the mismatch between $\hat h_e$ and $\hat h_m$
(the model waveform with the same physical parameters as $\hat h_e$)
will always be greater than the mismatch between $\hat h_e$ and $\hat
h_{\bar m}$.  So it is sufficient to require the maximum mismatch
parameter $\epsilon_{\max}$, that appears in
Eqs.~(\ref{e:DetectionLimit1}) and (\ref{e:DetectionAmpPhaseLimit}),
to have the maximum value allowed 
for $\epsilon_\mathrm{FF}$.  Thus $\epsilon_{\max}$
should have the value 0.005 for the current LIGO searches. 
It will be appropriate to revisit the issue of optimizing the 
values of  $\epsilon_\mathrm{MM}$ and $\epsilon_\mathrm{FF}$,
including the cost of producing the model waveforms versus 
the cost
of continually filtering with more populous template banks,
when fully numerical template banks are constructed in the future.

\subsection{Sufficient Conditions}
\label{s:SufficientConditionsIdeal}

The model waveform accuracy requirements for measurement,
Eq.~(\ref{e:MeasurementLimit3}), and detection,
Eq.~(\ref{e:DetectionAmpPhaseLimit}), are optimal in the sense that a
model waveform failing to meet these standards will cause a loss of
scientific information.  Conversely, a model waveform having smaller
errors than required will result in no added scientific value.
Unfortunately these accuracy requirements are somewhat complicated to
evaluate, since they place limits on the signal-weighted amplitude and
phase errors, $\overline{\delta\chi}$ and $\overline{\delta\Phi}$.
These weighted averages must be computed with the frequency-domain
waveform $h_m(f)$ and the detector noise spectrum $S_n(f)$.  While model
waveforms often scale in a trivial way with the total mass of the
gravitational wave source, the detector noise spectrum does not.  So
the model waveform errors must be evaluated separately for each mass,
and for each detector noise spectrum.  The purpose of this section is
to construct a set of simpler to apply accuracy requirements that are
nevertheless sufficient to guarantee that the optimal conditions
Eqs.~(\ref{e:MeasurementLimit3}) and (\ref{e:DetectionAmpPhaseLimit})
are satisfied.  While these sufficient conditions are stricter than
needed in many cases, we hope that their ease of use will allow the
waveform simulation community to employ them on a regular basis.  In
this section we present three different sufficient conditions that can
be applied to the frequency-domain representations of the waveforms,
and one condition that can be applied directly to the time-domain
waveforms.

The simplest sufficient conditions can be obtained by noting that
$\max |\delta\chi|\geq \overline{\delta\chi}$ and $\max |\delta \Phi|
\geq \overline{\delta\Phi}$, as a consequence of the definitions of
the signal-weighted waveform errors in
Eqs.~(\ref{e:AmplitudeAverageDef}) and (\ref{e:PhaseAverageDef}).  The
following are therefore sufficient conditions which ensure the optimal
waveform standard Eq.~(\ref{e:MeasurementLimit3}) is satisfied for 
measurement,
\begin{eqnarray}
\left(\max |\delta\chi|\right)^2 +
\left( \max |\delta { \Phi}|\right)^2 
&<& \frac{1}{\rho^2},
\label{e:MeasurementLimit4}
\end{eqnarray}
and Eq.~(\ref{e:DetectionAmpPhaseLimit}) is satisfied for detection
\begin{eqnarray}
\left(\max |\delta \chi|\right)^2 +
\left(\max |\delta { \Phi}|\right)^2 &<& 2\epsilon_{\max}.
\label{e:DetectionAmpPhaseMaxLimit}
\end{eqnarray}
We note that the maxima $\max|\delta\chi|$ and $\max |\delta \Phi|$ in
these limits refer to the frequency-domain waveform errors.  These
requirements are significantly simpler to apply than the optimal
waveform standards because they eliminate the detector noise spectrum
from the calculation, except for its contribution to the
signal-to-noise ratio $\rho$.  While simple to evaluate however, these
limits on the maxima are stronger than necessary, especially if the
amplitude or phase errors are sharply peaked in a narrow range of
frequencies or the maximum occurs at a frequency where the amplitude
of the wave is very small.  

A second, sometimes less demanding sufficient condition on the
accuracy of model waveforms can be obtained by noting that $\langle
\delta h| \delta h\rangle$ can be written as,
\begin{eqnarray}
\langle \delta h | \delta h\rangle &=& 
\langle \delta A | \delta A\rangle 
+\langle A_e \delta \Phi | A_e \delta \Phi\rangle,\nonumber\\
&\leq&\left(\frac{\max |\delta A|}{\bar n}\right)^2
+\left(\frac{\max|A_e \delta \Phi|}{\bar n}\right)^2, 
\label{e:MeasurementLimit5}
\end{eqnarray}
where $A_e=e^{\chi_e}$, $\delta A =A_e\delta\chi$,  and the
average detector noise $\bar n$ is defined as,
\begin{eqnarray}
\frac{1}{\bar n^{2}}\equiv 4\int_0^\infty \frac{df}{S_n(f)}=\langle 1|1\rangle.
\end{eqnarray}
We can use Eq.~(\ref{e:MeasurementLimit5}) to convert
Eqs.~(\ref{e:MeasurementLimit1}) and (\ref{e:DetectionLimit1}) into
alternate sufficient conditions for model waveform accuracy by
introducing a quantity $\bar C$:
\begin{eqnarray}
\bar C \equiv \frac{\rho \,\bar n}{A_e(f_\mathrm{c})}.
\label{e:barCdef}
\end{eqnarray}
In this expression $f_\mathrm{c}$ is a frequency that characterizes
the particular waveform.  For the equal mass binary black-hole
waveforms shown in
Fig.~\ref{f:FFThAmp} for example, a convenient choice might be
$f_\mathrm{c}=0.08/M$ which occurs near the point in the spectrum where
the two black holes merge.  This quantity $\bar C$ is the ratio of the
standard signal-to-noise measure $\rho$, and the non-standard measure
$A_e(f_\mathrm{c})/\bar n$.  Using the definition of $\bar C$ and
Eq.~(\ref{e:MeasurementLimit5}), we can convert
Eq.~(\ref{e:MeasurementLimit1}) into a simple sufficient condition on
the model waveform accuracy for measurement,
\begin{eqnarray}
\left[\frac{\max |\delta A|}{A_e(f_\mathrm{c})}\right]^2 +
\left[\frac{\max|A_e \delta \Phi|}{A_e(f_\mathrm{c})}\right]^2 &<&
\left(\frac{\bar C}{\rho}\right)^2,
\label{e:MeasurementLimit6}
\end{eqnarray}
and Eq.~(\ref{e:DetectionLimit1}) into a sufficient condition for 
detection purposes,
\begin{eqnarray}
\left[\frac{\max |\delta A|}{A_e(f_\mathrm{c})}\right]^2 
+ \left[\frac{\max|A_e
    \delta \Phi|}{A_e(f_\mathrm{c})}\right]^2 
&<& 2 \epsilon_{\max}\bar C^2.
\label{e:DetectionAmpPhaseMaxLimit2}
\end{eqnarray}
\noindent We note that the errors bounded in
Eqs.~(\ref{e:MeasurementLimit4}) and
(\ref{e:DetectionAmpPhaseMaxLimit}) are the logarithmic amplitude and
phase errors, while those in Eqs.~(\ref{e:MeasurementLimit6}) and
(\ref{e:DetectionAmpPhaseMaxLimit2}) are errors relative to the fixed
characteristic amplitude $A_e(f_\mathrm{c})$.  The limits in
Eqs.~(\ref{e:MeasurementLimit6}) and
(\ref{e:DetectionAmpPhaseMaxLimit2}) may therefore be more useful,
because they avoid the unnecessarily restrictive conditions on
$\max|\delta\chi|$ and $\max |\delta\Phi|$ when those maxima occur at
frequencies where the amplitude of the waveform is small.

The requirements in Eqs.~(\ref{e:MeasurementLimit6}) and
(\ref{e:DetectionAmpPhaseMaxLimit2}) have the disadvantage, however,
of involving the quantity $\bar C$ which depends on the details of the
waveform and the detector noise spectrum.  Nevertheless, this quantity
can easily be evaluated for a given class of waveforms and a
given detector noise spectrum.  For example, Fig.~\ref{f:CSNrat}
illustrates $\bar C$ for the case of non-spinning equal-mass binary
black-hole waveforms with $f_\mathrm{c}\approx
0.08/M$ using the Initial LIGO noise curve~\cite{InitialLIGONoise}  
with a 40 Hz low frequency cutoff~\cite{Abbott:2007kva}, 
and the Advanced LIGO noise curve~\cite{AdvancedLIGONoise} {with a
10 Hz low frequency cutoff. 
The scale factor $M_\odot$ used in this figure is the mass of the sun, 
$M_\odot=2.0\times 10^{33}$~g,
which in geometrical time units is $M_\odot = 4.9\times10^{-6}$~s.
This curve shows that $\bar C \gtrsim 0.6$ for the mass
range of non-spinning binary black-hole systems of primary 
relevance to Initial LIGO, $5\leq
M/M_\odot \leq 100$~\cite{LIGO-G080178-04-Z}.  
Thus for Initial LIGO it is sufficient to
enforce Eqs.~(\ref{e:MeasurementLimit6}) and
(\ref{e:DetectionAmpPhaseMaxLimit2}) for the case $\bar C \approx
0.6$.  The mass range for non-spinning 
binary black-hole systems of primary
relevance for Advanced LIGO extends to, $5\leq M/M_\odot \leq 400$, 
because the low frequency cutoff is 10 Hz instead of 40 Hz.
For Advanced LIGO then, it is appropriate to use the minimum value
$\bar C \approx 0.06$ when enforcing Eqs.~(\ref{e:MeasurementLimit6}) and
(\ref{e:DetectionAmpPhaseMaxLimit2}). 
\begin{figure}
\centerline{\includegraphics[width=3in]{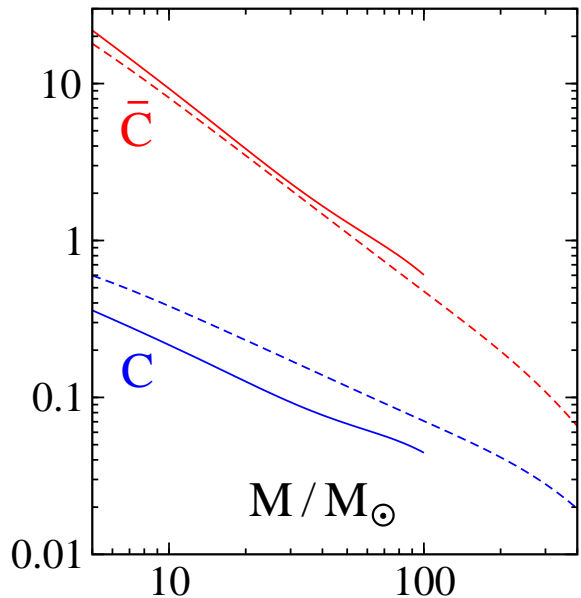}}
\caption{\label{f:CSNrat} 
Solid curves illustrate $\bar C$ and $C$, as defined in
Eqs.~(\ref{e:barCdef}) and (\ref{e:SNRatioRatio}), as functions of the
total mass for non-spinning equal-mass binary black-hole waveforms and
the initial LIGO noise spectrum.  Dashed curves give the same
quantities computed with the Advanced LIGO noise curve.}
\end{figure}

A third sufficient condition can be obtained by noting that,
\begin{eqnarray}
\langle\delta h|\delta h\rangle & = &
4\int_0^\infty \frac{|\delta h(f)|^2 df}{S_n(f)},\nonumber\\ 
&\leq & \frac{2||\delta h(f)||^2}{\mathrm{min}\, S_n(f)}
\label{e:frequency-domain-limit}
\end{eqnarray}
where $||\delta h(f)||$ is the $L^2$ norm of $\delta h(f)$, defined
as
\begin{eqnarray}
||\delta h(f)||^2=2\int_0^\infty |\delta h(f)|^2 df.
\end{eqnarray}
The inequality in Eq.~(\ref{e:frequency-domain-limit}) can be
converted to sufficient conditions for the optimal waveform
requirements by introducing the quantity $C$,
\begin{eqnarray} 
C^2=\frac{\rho^2\,\mathrm{min}\, S_n(f)}{2||h_e(f)||^2},
\label{e:SNRatioRatio}
\end{eqnarray}
the ratio of the standard signal-to-noise measure $\rho$ to another
non-standard measure $||h_e(f)|| / \sqrt{\mathrm{min\,}S_n(f)}$.
Using this definition it is straightforward to convert
Eq.~(\ref{e:frequency-domain-limit}) into sufficient conditions for
the optimal error requirements of Eq.~(\ref{e:MeasurementLimit1}) for
measurement,
\begin{eqnarray}
\frac{||\delta h(f)||^2}{||h_e(f)||^2} &<& \left(\frac{C}{\rho}\right)^2, 
\label{e:frequency-domain-measurement1}
\end{eqnarray}
and Eq.~(\ref{e:DetectionLimit1}) for detection
\begin{eqnarray}
\frac{||\delta h(f)||^2}{||h_e(f)||^2}&<& 2\epsilon_\mathrm{max}C^2.
\label{e:frequency-domain-detection1} 
\end{eqnarray}
As with the previous requirements, these can be
written in terms of the amplitude and phase of the frequency-domain
waveform:  for measurement,
\begin{eqnarray}
\frac{||\delta A(f)||^2+||A_e \delta \Phi(f)||^2}
{||A_e(f)||^2} &<& \left(\frac{C}{\rho}\right)^2, 
\label{e:frequency-domain-measurement2}
\end{eqnarray}
and for detection
\begin{eqnarray}
\frac{||\delta A(f)||^2+||A_e \delta \Phi(f)||^2}
{||A_e(f)||^2} &<& 2\epsilon_\mathrm{max}C^2.
\label{e:frequency-domain-detection2} 
\end{eqnarray}
These requirements depend only on the average value of the model
waveform error $||\delta h(f)||$, and so they are considerably weaker
than (and in this sense are superior to) the conditions in
Eqs.~(\ref{e:MeasurementLimit6}) and
(\ref{e:DetectionAmpPhaseMaxLimit2}).  Their biggest drawback is their
dependence on the waveform and noise spectrum through the quantity
$C$.  It is straightforward to show that $C\leq 1$ following an
argument similar to that which led to
Eq.~(\ref{e:frequency-domain-limit}).  However the exact value of $C$
will depend on the details of the model waveform, including in
particular the mass of the gravitational wave source.  But $C$ can 
easily be evaluated
for a given class of model waveforms and a given detector
noise spectrum.  For example, Fig.~\ref{f:CSNrat} shows a graph of $C$
as a function of the mass for non-spinning equal-mass binary
black-hole waveforms (evaluated with the Initial and the Advanced LIGO
noise curves).  This figure shows that taking $\min C\approx 0.02$ in
Eqs.~(\ref{e:frequency-domain-measurement1}) and
(\ref{e:frequency-domain-detection1}) is sufficient for the binary
black-hole mass range $5\leq M/M_\odot\leq 400$ of primary relevance
to Advanced LIGO, while taking $\min C\approx 0.04$ is sufficient for
the mass range $5\leq M/M_\odot \leq 100$ of primary relevance to
Initial LIGO~\cite{LIGO-G080178-04-Z}.  We note that the
functions $\bar C$ and $C$ illustrated in Fig.~\ref{f:CSNrat} apply,
strictly speaking, only to the waveform illustrated in
Figs.~\ref{f:FFThAmp} and \ref{f:FFThPhase}.  A more comprehensive
investigation will be required to determine how sensitive these
functions are to other model-waveform parameters (like the mass ratio
and the spins of the black holes) and to other effects (like the
frequency range of a given model waveform).

Our discussion up to this point has focused on the development of
accuracy standards for the frequency-domain representations of model
waveforms.  This approach simplifies the analysis, and is natural from
the LIGO data analysis perspective.  However, model waveforms must
often be computed in the time domain, e.g. by numerical relativity
simulations.  While time-domain waveforms can be converted to the
frequency domain, doing so is somewhat delicate and requires making
judicious choices in performing the needed Fourier transforms---like
choosing appropriate windowing functions.  Having time-domain versions
of the needed accuracy standards would therefore make it much easier
for the waveform simulation community to monitor and deliver waveforms
of the needed accuracy.  The third set of sufficient waveform accuracy
requirements, Eqs.~(\ref{e:frequency-domain-measurement1}) and
(\ref{e:frequency-domain-detection1}), can easily be converted to
conditions on the time-domain waveforms.  This can be done because the
$L^2$ norm of a time-domain waveform is identical to the $L^2$ norm of its
frequency-domain counterpart, e.g. $||\delta h(t)||=||\delta h(f)||$, by
Parseval's theorem.  Thus, the following are the corresponding time-domain
waveform accuracy standards for measurement,
\begin{eqnarray}
\frac{||\delta h(t)||}{||h_e(t)||} &<& \frac{C}{\rho}, 
\label{e:time-domain-measurement1}
\end{eqnarray}
and for detection
\begin{eqnarray}
\frac{||\delta h(t)||}{||h_e(t)||}&<& \sqrt{2\epsilon_\mathrm{max}}\,C.
\label{e:time-domain-detection1} 
\end{eqnarray}
It is not generally possible to decompose the real time-domain
waveform unambiguously into amplitude and phase.  Therefore it is not
possible to construct meaningful time-domain analogs of the amplitude
and phase limits given in Eqs.~(\ref{e:frequency-domain-measurement2})
and (\ref{e:frequency-domain-detection2}).

\section{Including Calibration Errors}
\label{s:IncludingCalibrationErrors}

In this section we consider the implications of having a detector
response function that is not known with absolute precision. First we
establish a little notation.  Let $v(f)$ denote the direct
electronic output of the detector, and $R(f)$ the response function
that converts the raw output $v(f)$ into the inferred gravitational
wave signal $h(f)$:
\begin{eqnarray}
h(f)=R(f)v(f).
\end{eqnarray}
Let us assume that the measured response function $R(f)$ differs from
the correct exact response function $R_e(f)$ by 
$\delta R(f) = R(f)-R_e(f)$.  This
error in the response function will affect measurements in two ways.
First, the response of the instrument to a gravitational wave signal
$h_e$ will produce an electronic output $v_e$.  Using the measured
response function $R$, the signal will be interpreted as the waveform
$h=Rv_e=h_ee^{\delta\chi_R+i\delta\Phi_R}$, where the logarithmic
response function amplitude $\delta \chi_R$ and phase $\delta\Phi_R$
errors are defined by
\begin{eqnarray}
R= R_e+\delta R = R_e e^{\delta\chi_R+i\delta\Phi_R}.
\end{eqnarray}
Thus there will be a waveform error, 
\begin{eqnarray}
\delta
h_R=h_ee^{\delta\chi_R+i\delta\Phi_R}-h_e, 
\end{eqnarray}
caused by calibration errors in the instrument.  The second effect of
calibration error on measurements made with the instrument will be
errors in our knowledge of the characteristics of the noise in the
detector.  In particular the estimated power spectral density of the
noise $S_n$ will differ from the exact $S_e$ due to the calibration
error $\delta R$.  The estimated power spectral density $S_n$
will be related to $S_e$ by
\begin{eqnarray}
S_n(f) = S_e(f)e^{2\delta\chi_R}.
\label{e:MeasuredSh}
\end{eqnarray}

The idea now is to evaluate the effects of errors in the model
waveform, $h_m=h_ee^{\delta\chi_m+i\delta\Phi_m}$, plus the effects of
errors in the detector response function
$R=R_ee^{\delta\chi_R+i\delta\Phi_R}$, on the signal-to-noise ratio of
a detected gravitational wave signal:
\begin{eqnarray}
\rho_m=\frac{\langle h_e+\delta h_R|h_m\rangle}{\langle h_m|h_m\rangle^{1/2}},
\label{e:MeasuredSN}
\end{eqnarray}
where the inner product is evaluated with respect to the
estimated power spectral noise density
$S_n(f)$ of Eq.~(\ref{e:MeasuredSh}).  The needed calculation is quite
long, so we provide a few intermediate steps.  Keeping terms through
second order in the two types of errors, we find:
\begin{eqnarray}
&&\langle h_e+\delta h_R|h_m\rangle = \rho^2 
+\rho^2 \langle(\delta\chi_m-\delta\chi_R)\hat h_e|\hat h_e\rangle\nonumber\\
&&\qquad\qquad
+\frac{\rho^2}{2}\langle(\delta\chi_m-\delta\chi_R)\hat h_e|
(\delta\chi_m-\delta\chi_R)\hat h_e\rangle\nonumber\\
&&\qquad\qquad
-\frac{\rho^2}{2}\langle(\delta\Phi_m-\delta\Phi_R)\hat h_e|
(\delta\Phi_m-\delta\Phi_R)\hat h_e\rangle,\qquad\\
&&\langle h_m|h_m\rangle^{-1/2}=\frac{1}{\rho}
-\frac{1}{\rho}\langle(\delta\chi_m-\delta\chi_R)\hat h_e|\hat h_e\rangle
\nonumber\\
&&\qquad\qquad
-\frac{1}{\rho}\langle(\delta\chi_m-\delta\chi_R)\hat h_e|
(\delta\chi_m-\delta\chi_R)\hat h_e\rangle\nonumber\\
&&\qquad\qquad
+\frac{3}{2\rho}\langle(\delta\chi_m-\delta\chi_R)\hat h_e|\hat h_e\rangle^2.
\end{eqnarray}
Combining these using Eq.~(\ref{e:MeasuredSN}) gives an expression for
the effects of model and calibration errors on the measured
signal-to-noise ratio $\rho_m$:\footnote{
A special case of this expression for the
signal-to-noise ratio including the first-order 
calibration-error terms
was obtained previously by Bruce Allen~\cite{Allen1996}, 
and an expression including the 
second-order calibration
error terms (and some of the first-order model-error terms) 
was obtained previously 
by Sukanta Bose~\cite{Bose2005}.}}
\begin{eqnarray}
\rho_m&=& \rho
-\frac{\rho}{2}\langle(\delta\chi_m-\delta\chi_R)\hat h_e|
(\delta\chi_m-\delta\chi_R)\hat h_e\rangle\nonumber\\
&&-\frac{\rho}{2}\langle(\delta\Phi_m-\delta\Phi_R)\hat h_e|
(\delta\Phi_m-\delta\Phi_R)\hat h_e\rangle\nonumber\\
&&+\frac{\rho}{2}\langle(\delta\chi_m-\delta\chi_R)\hat h_e|\hat h_e\rangle^2.
\end{eqnarray}

It is illuminating to write this expression in the form,
\begin{eqnarray}
\rho_m = \rho-\frac{1}{2\rho}\langle(\delta h_m-\delta h_R)_\perp|
(\delta h_m-\delta h_R)_\perp\rangle,
\label{e:modelpluscalibrationerror}
\end{eqnarray}
where
\begin{eqnarray}
(\delta h_m - \delta h_R)_\perp =
\delta h_m -\delta h_R 
- \hat h_e \langle \delta h_m - \delta h_R|\hat h_e\rangle.
\label{e:delta_h_perp_def}
\end{eqnarray}
The dependence of the measured signal-to-noise ratio $\rho_m$ on the
waveform errors, $\delta h_m$ and $\delta h_R$, can be understood as
follows: If (hypothetically) 
the modeling errors were identical to the calibration
errors, the measured signal and template would still be identical, so
the measured matched-filtering signal-to-noise ratio would be unchanged.
If (again hypothetically) 
the net waveform error, $\delta h_m-\delta h_R$, were proportional
to the exact waveform, $h_e$, the measured signal-to-noise ratio
$\rho_m$ would be unchanged because such errors merely re-scale the
template, which has no effect on $\rho_m$.  Thus only net waveform
errors that are orthogonal to $h_e$, in the sense of
Eq.~(\ref{e:delta_h_perp_def}), actually 
contribute to losses in $\rho_m$.

We can use this basic expression,
Eq.~(\ref{e:modelpluscalibrationerror}), for the combined effects of
model waveform error and calibration error to derive a useful
inequality on the signal-to-noise ratio:
\begin{eqnarray}
\rho_m &\geq& \rho-\frac{1}{2\rho}\langle\delta h_m-\delta h_R|
\delta h_m-\delta h_R\rangle,\nonumber\\
&\geq&\rho-\frac{1}{2\rho}\left[\sqrt{\langle\delta h_m|\delta h_m\rangle}
+\sqrt{\langle\delta h_R|\delta h_R\rangle}\right]^2.\qquad
\label{e:calibrationlimit0}
\end{eqnarray}
To aid our understanding of this expression, it is  useful to
define the ratio of the model waveform error to the response function
error, $\eta$:
\begin{eqnarray}
\langle\delta h_m|\delta h_m\rangle=\eta^2
\langle\delta h_R|\delta h_R\rangle,
\end{eqnarray}
which allows us to re-write Eq.~(\ref{e:calibrationlimit0}) as
\begin{eqnarray}
\rho_m\geq \rho -\frac{1}{2\rho}(1+\eta)^2
\langle\delta h_R|\delta h_R\rangle.
\label{e:calibrationSNlimit0}
\end{eqnarray}

If $\eta$ becomes too small, then the error in the measured
signal-to-noise ratio $\rho_m$ is dominated by calibration errors, and
further reductions in the model waveform error have little effect
on our ability to make measurements or detections.
The idea then is to place a limit on $\eta$ which ensures the model waveform
is only as accurate as it needs to be to achieve the ideal-detector
accuracy standards of Sec.~\ref{s:IdealDetectorCase}.  It is
appropriate therefore to require that $\eta$ be no smaller than some minimum
cutoff: $\eta\geq\eta_{\min}$.  In other words, the model waveform
error need never be made smaller than 
\begin{eqnarray}
\langle\delta h_m|\delta h_m\rangle\geq\eta_{\min}^2
\langle\delta h_R|\delta h_R\rangle.
\label{e:calibrationlimit}
\end{eqnarray}
This inequality can also be expressed as a condition on the
signal-weighted amplitude and phase errors defined in
Sec.~\ref{s:IdealDetectorCase}:
\begin{eqnarray}
\overline{\delta\chi}_m^{\,2}+\overline{\delta\Phi}_m^{\,2}\geq \eta_{\min}^2
\left(\overline{\delta\chi}_R^{\,2}+\overline{\delta\Phi}_R^{\,2}\right).
\label{e:calibrationlimit1}
\end{eqnarray}
These lower limits, Eqs.~(\ref{e:calibrationlimit}) or
(\ref{e:calibrationlimit1}), on the model waveform error must be
imposed simultaneously with the appropriate upper limits derived in
Sec.~\ref{s:IdealDetectorCase} for measurement,
Eq.~(\ref{e:MeasurementLimit3}), or detection,
Eq.~(\ref{e:DetectionAmpPhaseLimit}).  Calibration error will
interfere with the operation of a detector whenever it is impossible
to satisfy the ideal-detector accuracy requirements of 
Sec.~\ref{s:IdealDetectorCase} and the calibration-error
lower limits simultaneously. 

To determine the appropriate value for $\eta_{\min}$, we 
note from Eq.~(\ref{e:calibrationSNlimit0}) that a model waveform having
the minimal error $\eta=\eta_{\min}$ could have an effect
on the measured signal-to-noise ratio that is as large as, 
\begin{eqnarray}
\rho_m\geq \rho -\frac{1}{2\rho}(1+\eta_{\min})^2
\langle\delta h_R|\delta h_R\rangle.
\label{e:calibrationSNlimit}
\end{eqnarray}
Setting $\eta_{\min}=1$ corresponds to one natural choice: making the
model waveform errors greater  
than or equal to the calibration errors.
However, we see from Eq.~(\ref{e:calibrationSNlimit}) that the
signal-to-noise ratio may be degraded by up to four times the effect
of calibration error alone in this case.  So this obvious
choice for $\eta_{\min}$ is probably too large.

Given the relatively low cost of producing improved model
waveforms (compared to the cost of improving the hardware needed to
reduce calibration errors), it makes sense to adopt a stricter
standard for $\eta_{\min}$.  Another natural choice is to
require that $\eta_{\min}$ be small enough to ensure that
the most accurate model waveforms have no larger effect on the
signal-to-noise ratio than calibration errors alone.  From
Eq.~(\ref{e:calibrationSNlimit}) we see that this requires
$(1+\eta_{\min})^2\leq 2$ or equivalently $\eta_{\min}\lesssim
0.4$.  Reducing $\eta_{\min}$ below this value has little
effect on the signal-to-noise ratio; while quickly increasing the
computational cost of the most accurate 
model waveforms.  So this choice should be
close to optimal.

The most recent publicly available calibration data from
Initial LIGO (S4), see Figs.\ 23 and 24 of Ref.~\cite{LIGOS4Calibration},
has the following limits on the frequency-domain calibration error:
$0.03\lesssim\sqrt{\delta\chi_R^2+\delta\Phi_R^2}\lesssim 0.09$
for the L1 detector and
$0.06\lesssim\sqrt{\delta\chi_R^2+\delta\Phi_R^2}\lesssim 0.12$
for the H1 detector.  Therefore the appropriate minimum error
requirement, including the effects of calibration
errors from Eq.~(\ref{e:calibrationlimit1}), for Initial LIGO
is\footnote{Although the calibration of the final data from 
Initial LIGO (S5) will be different from that of S4, the accuracy of 
the calibration is expected to be similar.}
\begin{eqnarray}
\sqrt{\overline{\delta\chi}_m^{\,2}+\overline{\delta\Phi}_m^{\,2}}
&\geq&\eta_{\min}
\sqrt{(\min|\delta\chi_R|)^{\,2}+(\min|\delta\Phi_R|)^{\,2}}
\nonumber\\
&\gtrsim&
0.012.
\label{e:CalibrationSufficient}
\end{eqnarray}
Model waveform errors smaller than this limit will always be dominated by
calibration errors.

The magnitude of the calibration error in Initial LIGO is small enough
that the calibration requirement, 
Eq.~(\ref{e:CalibrationSufficient}), is consistent
with our basic measurement requirements,
Eq.~(\ref{e:MeasurementLimit3}) or (\ref{e:time-domain-measurement1}),
for all but the very strongest sources: $\rho\gtrsim 80$.  Therefore
the presence of calibration error will not affect the ideal-detector
waveform accuracy standards for LIGO measurements, unless an extremely
strong source is detected.  This is not to say that the presence of
calibration error will have no effect on the accuracy of measurements
for weaker sources; but the question of exactly how large
those calibration error effects are in that case 
must be decided by a somewhat more
detailed analysis than the one presented here.

The lower limit on model waveform error that arises from the presence
of calibration error is always consistent with the condition on the
waveform error needed for detection,
Eq.~(\ref{e:DetectionAmpPhaseLimit}).  From
Eq.~(\ref{e:calibrationSNlimit}), the fractional change in the
signal-to-noise ratio caused by calibration error is
\begin{eqnarray}
\left|\frac{\delta \rho}{\rho}\right| \leq (1+\eta_\mathrm{min})^2
\frac{\langle\delta h_R|\delta h_R\rangle}{2\rho^2}.
\label{e:fractionalSNlimit}
\end{eqnarray}
For Initial LIGO 
$\langle\delta h_R|\delta h_R\rangle\lesssim 0.014$ 
(from the calibration error measurements quoted above)
thus the signal-to-noise ratio $\rho$ would have to be smaller than
about 1.7 (which is below any reasonable statistical threshold
for detection) before the right side of Eq.~(\ref{e:fractionalSNlimit})
exceeds the detection limit $\epsilon_{\max}=0.005$.
Thus, the current level of 
LIGO calibration errors 
are never likely to influence the waveform accuracy requirements established in
Sec.~\ref{s:SignalDetectionIdeal} for detection.

\section{Discussion}
\label{s:Discussion}

In this paper we have developed a set of accuracy standards for model
gravitational waveforms.  These standards are designed to ensure that
model waveforms are accurate enough to support the parameter
measurement and detection needs of the gravitational wave data
analysis community---without compromising the scientific content of
the data, and without placing needless demands for accuracy on the
waveform modeling community.  In Sec.~\ref{s:IdealDetectorCase} we
developed an optimal requirement for measurement purposes in
Eq.~(\ref{e:MeasurementLimit3}) and for detection purposes in
Eq.~(\ref{e:DetectionAmpPhaseLimit}).  These optimal standards place
limits on the signal averaged amplitude and phase errors,
$\overline{\delta\chi}$ and $\overline{\delta\Phi}$ respectively, as
defined in Eqs.~(\ref{e:AmplitudeAverageDef}) and
(\ref{e:PhaseAverageDef}).  The first row of Table~\ref{t:TableBasic}
summarizes these accuracy standards (assuming the amplitude and phase
errors to be comparable): $\rho$ represents the standard
signal-to-noise ratio of the waveform, and $\epsilon_{\max}$
represents the maximum signal-to-noise mismatch tolerated by a given
detection procedure.
\begin{table}
\begin{center}
\begin{tabular}{|c|c|c|c|}\hline
Waveform Error& Equation    &Measurement & Detection \\
Diagnostic&Numbers &Requirement&Requirement\\
\hline
$\overline{\delta\Phi}$ &\ref{e:MeasurementLimit3},
                         \ref{e:DetectionAmpPhaseLimit}
                            & $1/\sqrt{2}\,\rho$ 
                            & $\sqrt{\epsilon_{\max}}$ \\
$\max | \delta\Phi|$    &\ref{e:MeasurementLimit4},
                         \ref{e:DetectionAmpPhaseMaxLimit}
                            & $1/\sqrt{2}\,\rho$ 
                            & $\sqrt{\epsilon_{\max}}$ \\
$\max | A_e\delta\Phi|/A_e(f_\mathrm{c})$  
                        &\ref{e:MeasurementLimit6},
                         \ref{e:DetectionAmpPhaseMaxLimit2}
                            & $\bar C/\sqrt{2}\,\rho$ 
                            & $\sqrt{\epsilon_{\max}}\,\bar C $ \\
$|| A_e\delta\Phi(f)||/||A_e(f)||$  
                        &\ref{e:frequency-domain-measurement2},
                         \ref{e:frequency-domain-detection2}
                            & $C/\sqrt{2}\,\rho$ 
                            & $\sqrt{\epsilon_{\max}}\,C$ \\
$||\delta h(t)||/||h_e(t)||$  &\ref{e:time-domain-measurement1},
                                \ref{e:time-domain-detection1}
                                  & $C/\rho$ 
                                  & $\sqrt{2\epsilon_{\max}}\,C$ \\
\hline
\end{tabular}
\end{center}
\caption{Summary of model waveform accuracy requirements for various
waveform error diagnostics for measurement purposes (column three) and for
detection purposes (column four).}
\label{t:TableBasic}
\end{table}

\begin{figure}
\centerline{\includegraphics[width=3in]{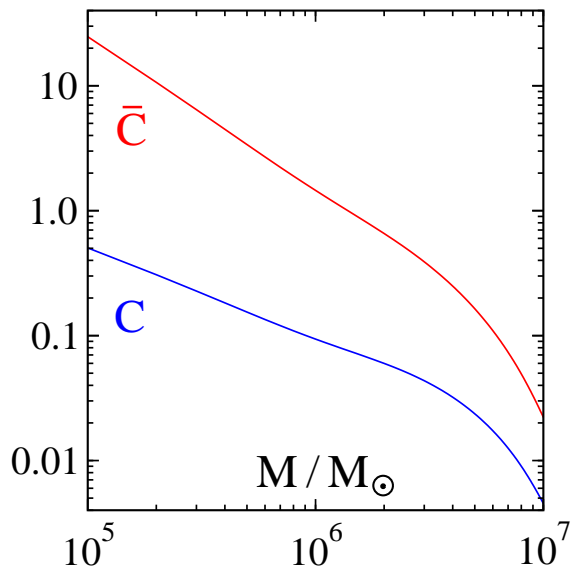}}
\caption{\label{f:CSNrat_LISA} Curves illustrate $\bar C$ and
$C$, as defined in Eqs.~(\ref{e:barCdef}) and
(\ref{e:SNRatioRatio}), as functions of the total mass for
non-spinning equal-mass binary black-hole waveforms using
the  Barack and Cutler~\cite{Barack2004} approximate
LISA noise spectrum.}
\end{figure}

Also listed in Table~\ref{t:TableBasic} are summaries of several
sufficient conditions on the waveform accuracy developed in
Sec.~\ref{s:SufficientConditionsIdeal} and described in detail in
Eqs.~(\ref{e:MeasurementLimit4}), (\ref{e:DetectionAmpPhaseMaxLimit}),
(\ref{e:MeasurementLimit6}), (\ref{e:DetectionAmpPhaseMaxLimit2}),
(\ref{e:frequency-domain-measurement2}),
(\ref{e:frequency-domain-detection2}),
(\ref{e:time-domain-measurement1}), and
(\ref{e:time-domain-detection1}).
The table entries for phase errors
assume that the amplitude and phase errors are comparable.
These sufficient conditions are
somewhat stronger than needed, but if satisfied they ensure the
optimal requirements are satisfied as well.  They are much simpler
to apply.  The quantities $\bar C$ and $C$ which appear in some of
these conditions, defined in Eqs.~(\ref{e:barCdef}) and
(\ref{e:SNRatioRatio}), compare different signal-to-noise measures of
the waveforms.  These quantities, which are waveform and detector noise
dependent, are illustrated for non-spinning equal-mass binary
black-hole waveforms in Figs.~\ref{f:CSNrat} and \ref{f:CSNrat_LISA}
for the LIGO and LISA detectors, respectively.

To apply these waveform accuracy standards to a particular detector,
we must know the values of the parameters that appear in the
requirements summarized in Table~\ref{t:TableBasic}.  In particular we
need to know $\max \rho$, $\epsilon_{\max}$, $\min \bar C$ and $\min
C$ for that detector.  The maximum signal-to-noise ratio, $\max
\rho$, expected for Advanced LIGO is rather difficult to estimate,
since the numbers and distribution of the relevant sources are not
known yet.  Here we need an estimate for $\max \rho$ only to compute
the appropriate Advanced LIGO values of the requirements in
Table~\ref{t:TableBasic}, so we take a rather simplistic approach: The
threshold signal-to-noise ratio for detecting compact binary inspirals
currently used by Initial LIGO is about $\rho\approx
9$ (for S4, the most recent data published)~\cite{Abbott:2007xi}.  
Advanced LIGO is expected to be about ten
(or perhaps as much as fifteen) 
times more sensitive than Initial LIGO. Therefore, if no binary
black-hole system is observed in the Initial LIGO (S5) data, then it
is likely that the maximum signal-to-noise ratio for such events in
the first year or two of Advanced LIGO observations will be no larger
than about $\max \rho\approx 100$. (This estimate might need to be 
increased or decreased by about 50\% depending on the final sensitivity
improvement of Advanced LIGO, and the final threshold adopted for
detections in the S5 data; this final threshold will
depend on the number of non-Gaussian artefacts 
found in the data.)  
Of course, if a binary black-hole signal
is detected by Initial LIGO, then the expectation would be that a
similar signal with ten to fifteen times that signal-to-noise ratio will be
detected by Advanced LIGO.  If Advanced LIGO template banks are
constructed in the same way as those for Initial LIGO, then
$\epsilon_{\max}=0.005$ is the appropriate mismatch tolerance for
Advanced LIGO as well.  We also need estimates of the quantities $\min
\bar C$ and $\min C$ for some of the waveform accuracy requirements.
From Fig.~\ref{f:CSNrat} we see that the appropriate choices for these
quantities for binary black-hole systems with masses in the range
$5\leq M/M_\odot \leq 400$ for the Advanced LIGO detector are $\min
\bar C\approx 0.06$ and $\min C\approx 0.02$.  Using these parameter
estimates, we summarize in Table~\ref{t:TableLIGO} our current
expectations for the model waveform accuracy standards that will be
needed for Advanced LIGO data analysis.
\begin{table}
\begin{center}
\begin{tabular}{|c|c|c|c|}\hline
Waveform Error&Equation&Measurement & Detection \\
Diagnostic&Numbers&Requirement&Requirement\\
\hline
$\overline{\delta\Phi}$ &\ref{e:MeasurementLimit3},
                         \ref{e:DetectionAmpPhaseLimit}
                            & 0.007 
                            & 0.07  \\
$\max | \delta\Phi|$    &\ref{e:MeasurementLimit4},
                         \ref{e:DetectionAmpPhaseMaxLimit}
                            & 0.007  
                            & 0.07  \\
$\max | A_e\delta\Phi|/A_e(f_\mathrm{c})$  
                        &\ref{e:MeasurementLimit6},
                         \ref{e:DetectionAmpPhaseMaxLimit2}
                            & 0.0004   
                            & 0.004  \\
$|| A_e\delta\Phi(f)||/||A_e(f)||$  
                        &\ref{e:frequency-domain-measurement2},
                         \ref{e:frequency-domain-detection2}
                            & 0.00014  
                            & 0.0014  \\
$||\delta h(t)||/||h_e(t)||$  &\ref{e:time-domain-measurement1},
                                \ref{e:time-domain-detection1}
                                  & 0.0002 
                                  & 0.002 \\
\hline
\end{tabular}
\end{center}
\caption{Summary of model waveform accuracy requirements for the
Advanced LIGO detector using various waveform error diagnostics for
measurement purposes (column three) and for detection purposes (column
four).}
\label{t:TableLIGO}
\end{table}

The detection requirements listed in the first two rows of
Table~\ref{t:TableLIGO} should
apply equally to Initial and to Advanced LIGO, since these
requirements depends only on the parameter $\epsilon_{\max}$ which
is determined by the properties of the search template bank and
the accepted level of missed detections rather than the sensitivity of
the detector.  The somewhat stronger sufficient conditions that
appear in rows 3-5 of Table~\ref{t:TableLIGO} also depend on the
constants $\bar C$ and $C$, which are larger for the Initial
LIGO case by factors of 10 and 2 respectively because the appropriate
mass range for Initial LIGO is smaller.  
The appropriate Initial LIGO requirements for
measurement are less clear.  The Initial LIGO detector is about ten
times less sensitive than the planned Advanced LIGO detector, so the
expectation is that the accuracy requirements should be about ten
times weaker than those listed in Table~\ref{t:TableLIGO}.  However
until a detection is actually made, no measurement requirements will
be needed at all.

\begin{table}
\begin{center}
\begin{tabular}{|c|c|c|c|}\hline
Waveform Error&Equation&Measurement & Detection \\
Diagnostic&Numbers &Requirement&Requirement\\
\hline
$\overline{\delta\Phi}$ &\ref{e:MeasurementLimit3},
                         \ref{e:DetectionAmpPhaseLimit}
                            & $2\cdot 10^{-4}$ 
                            & 0.07  \\
$\max | \delta\Phi|$    &\ref{e:MeasurementLimit4},
                         \ref{e:DetectionAmpPhaseMaxLimit}
                            & $2\cdot 10^{-4}$  
                            & 0.07  \\
$\max | A_e\delta\Phi|/A_e(f_\mathrm{c})$  
                        &\ref{e:MeasurementLimit6},
                         \ref{e:DetectionAmpPhaseMaxLimit2}
                            & $4\cdot 10^{-6}$   
                            & 0.001  \\
$|| A_e\delta\Phi(f)||/||A_e(f)||$  
                        &\ref{e:frequency-domain-measurement2},
                         \ref{e:frequency-domain-detection2}
                            & $7\cdot 10^{-7}$  
                            & 0.0003  \\
$||\delta h(t)||/||h_e(t)||$  &\ref{e:time-domain-measurement1},
                                \ref{e:time-domain-detection1}
                                  & $1\cdot10^{-6}$
                                  & 0.0004 \\
\hline
\end{tabular}
\end{center}
\caption{Summary of model waveform accuracy requirements for the
LISA detector for various waveform error diagnostics for
measurement purposes (column three) and for detection purposes (column
four).}
\label{t:TableLISA}
\end{table}
While the timetable and the technical specifications of the planned LISA
detector are still being developed, we think it is appropriate
to consider here what waveform accuracy standards this mission will
eventually require from the waveform simulation community.  The
maximum signal-to-noise ratio for supermassive binary black-hole
observations by LISA is expected to be much larger than the stellar
mass black-hole observations made by LIGO.  Supermassive black-hole
mergers are expected to occur at a rate of about one merger per year
within a sphere that extends to cosmological redshift
$z=2$~\cite{Sesana:2005qn,Sesana:2005}.  The largest signal-to-noise ratio for
non-spinning equal-mass binaries located at 
15 Gpc (redshift $z\approx 2$)
is about $\max\rho \approx 4\times 10^3$ for an optimally oriented
system with total mass $5\times
10^6M_\odot$~\cite{Buonanno-Cook-Pretorius:2007}.  Thus the
requirement on the accuracy of model waveforms for measurement
purposes is likely to be much more demanding for LISA than for LIGO.  If the
LISA template banks are constructed in the same way as the LIGO
template banks, then it is appropriate to assume the maximum mismatch
parameter is $\epsilon_{\max}=0.005$ as in the LIGO case.
Figure~\ref{f:CSNrat_LISA} illustrates the quantities $\bar C$ and $C$
based on the approximate LISA noise curve constructed by Barack and
Cutler~\cite{Barack2004} (cf. their Eq. 30) using a $10^{-4}$ Hz low
frequency cutoff.  It follows that $\min
\bar C \approx 0.02$ and $\min C \approx 0.004$ for binary systems
with total masses in the range $10^5 < M/M_\odot < 10^7$.  Using these
parameter estimates, we summarize in Table~\ref{t:TableLISA} our
current expectations for the model waveform accuracy standards that
will be needed for LISA data analysis.

The accuracy requirements for measurement and detection that we
discuss here provide upper limits on the allowed errors of model
waveforms.  These upper limits are the only requirements on model
waveform accuracy when the detector is ideal, i.e. when the response
function of the detector is known with absolute precision.  In
Sec.~\ref{s:IncludingCalibrationErrors} we discuss the additional
requirements that must be imposed when the response function has
errors.  We show that the model waveform error need never be decreased
to a level below a certain fraction, $\eta_{\min}\approx 0.4$, of the
response function error.  For Initial LIGO the calibration error is
small enough that it will not affect the ability of the instrument to
make detections at all.  The effect of this error on the ability of
Initial LIGO to make measurements is more complicated: Calibration
errors will degrade the quality of measurements made on sources with
large signal to noise ratios, $\rho\gtrsim 80$, and decreasing the
model waveform error below $\eta_{\min}$ times the response function
error will not improve these measurements substantially.  For weaker
signals, $\rho\lesssim 80$, the calibration error is likely to degrade
the quality of measurements to some extent, but the ideal-detector
model-waveform accuracy standards should nevertheless be enforced in
this case.

\acknowledgments We thank Emanuele Berti, Curt Cutler, and Michele
Vallisneri for illuminating discussions on the subject of
gravitational wave data analysis, and its particular application to LISA.
We also thank Stephen Fairhurst, Kip Thorne, and Alan Weinstein
for a number of useful comments on a preliminary draft of this paper.
This research was supported in part by a grant to Caltech from the
Sherman Fairchild Foundation, by NSF grants DMS-0553302,
PHY-0601459, and PHY-0652995 to Caltech, 
by NSF grant PHY-0555628 to Penn
State, and by the Penn State Center for Gravitational Wave Physics under NSF
cooperative agreement PHY-0114375.

\bibliography{../References/References}

\begin{thebibliography}{39}
\expandafter\ifx\csname natexlab\endcsname\relax\def\natexlab#1{#1}\fi
\expandafter\ifx\csname bibnamefont\endcsname\relax
  \def\bibnamefont#1{#1}\fi
\expandafter\ifx\csname bibfnamefont\endcsname\relax
  \def\bibfnamefont#1{#1}\fi
\expandafter\ifx\csname citenamefont\endcsname\relax
  \def\citenamefont#1{#1}\fi
\expandafter\ifx\csname url\endcsname\relax
  \def\url#1{\texttt{#1}}\fi
\expandafter\ifx\csname urlprefix\endcsname\relax\def\urlprefix{URL }\fi
\providecommand{\bibinfo}[2]{#2}
\providecommand{\eprint}[2][]{\url{#2}}

\bibitem[{\citenamefont{Abbott et~al.}(2007)}]{Abbott:2007kva}
\bibinfo{author}{\bibfnamefont{B.}~\bibnamefont{Abbott}} \bibnamefont{et~al.}
  (\bibinfo{collaboration}{LIGO Scientific Collaboration})
  (\bibinfo{year}{2007}), \eprint{arXiv:0711.3041}.

\bibitem[{\citenamefont{Acernese et~al.}(2008)}]{Acernese:2008zz}
\bibinfo{author}{\bibfnamefont{F.}~\bibnamefont{Acernese}} \bibnamefont{et~al.}
  (\bibinfo{collaboration}{Virgo Collaboration}), \bibinfo{journal}{Class.
  Quant. Grav.} \textbf{\bibinfo{volume}{25}}, \bibinfo{pages}{114045}
  (\bibinfo{year}{2008}).

\bibitem[{\citenamefont{Grote}(2008)}]{Grote:2008zz}
\bibinfo{author}{\bibfnamefont{H.}~\bibnamefont{Grote}}
  (\bibinfo{collaboration}{GEO600 Collaboration}), \bibinfo{journal}{Class.
  Quant. Grav.} \textbf{\bibinfo{volume}{25}}, \bibinfo{pages}{114043}
  (\bibinfo{year}{2008}).

\bibitem[{\citenamefont{Shibata and Uryu}(2002)}]{Shibata2002}
\bibinfo{author}{\bibfnamefont{M.}~\bibnamefont{Shibata}} \bibnamefont{and}
  \bibinfo{author}{\bibfnamefont{K.}~\bibnamefont{Uryu}},
  \bibinfo{journal}{Prog. Theor. Phys.} \textbf{\bibinfo{volume}{107}},
  \bibinfo{pages}{265} (\bibinfo{year}{2002}).

\bibitem[{\citenamefont{Pretorius}(2005)}]{Pretorius2005a}
\bibinfo{author}{\bibfnamefont{F.}~\bibnamefont{Pretorius}},
  \bibinfo{journal}{Phys.\ Rev.\ Lett.} \textbf{\bibinfo{volume}{95}},
  \bibinfo{eid}{121101} (\bibinfo{year}{2005}).

\bibitem[{\citenamefont{Campanelli et~al.}(2006)\citenamefont{Campanelli,
  Lousto, Marronetti, and Zlochower}}]{Campanelli2006a}
\bibinfo{author}{\bibfnamefont{M.}~\bibnamefont{Campanelli}},
  \bibinfo{author}{\bibfnamefont{C.~O.} \bibnamefont{Lousto}},
  \bibinfo{author}{\bibfnamefont{P.}~\bibnamefont{Marronetti}},
  \bibnamefont{and}
  \bibinfo{author}{\bibfnamefont{Y.}~\bibnamefont{Zlochower}},
  \bibinfo{journal}{Phys.\ Rev.\ Lett.} \textbf{\bibinfo{volume}{96}},
  \bibinfo{eid}{111101} (\bibinfo{year}{2006}).

\bibitem[{\citenamefont{Faber et~al.}(2006)\citenamefont{Faber, Baumgarte,
  Shapiro, and Taniguchi}}]{Faber2006}
\bibinfo{author}{\bibfnamefont{J.~A.} \bibnamefont{Faber}},
  \bibinfo{author}{\bibfnamefont{T.~W.} \bibnamefont{Baumgarte}},
  \bibinfo{author}{\bibfnamefont{S.~L.} \bibnamefont{Shapiro}},
  \bibnamefont{and}
  \bibinfo{author}{\bibfnamefont{K.}~\bibnamefont{Taniguchi}},
  \bibinfo{journal}{Astrophys. J.} \textbf{\bibinfo{volume}{641}},
  \bibinfo{pages}{L93} (\bibinfo{year}{2006}).

\bibitem[{\citenamefont{Baker et~al.}(2006)\citenamefont{Baker, Centrella,
  Choi, Koppitz, and van Meter}}]{Baker2006a}
\bibinfo{author}{\bibfnamefont{J.~G.} \bibnamefont{Baker}},
  \bibinfo{author}{\bibfnamefont{J.}~\bibnamefont{Centrella}},
  \bibinfo{author}{\bibfnamefont{D.-I.} \bibnamefont{Choi}},
  \bibinfo{author}{\bibfnamefont{M.}~\bibnamefont{Koppitz}}, \bibnamefont{and}
  \bibinfo{author}{\bibfnamefont{J.}~\bibnamefont{van Meter}},
  \bibinfo{journal}{Phys.\ Rev.\ Lett.} \textbf{\bibinfo{volume}{96}},
  \bibinfo{eid}{111102} (\bibinfo{year}{2006}).

\bibitem[{\citenamefont{Diener et~al.}(2006)\citenamefont{Diener, Herrmann,
  Pollney, Schnetter, Seidel, Takahashi, Thornburg, and
  Ventrella}}]{Diener2006}
\bibinfo{author}{\bibfnamefont{P.}~\bibnamefont{Diener}},
  \bibinfo{author}{\bibfnamefont{F.}~\bibnamefont{Herrmann}},
  \bibinfo{author}{\bibfnamefont{D.}~\bibnamefont{Pollney}},
  \bibinfo{author}{\bibfnamefont{E.}~\bibnamefont{Schnetter}},
  \bibinfo{author}{\bibfnamefont{E.}~\bibnamefont{Seidel}},
  \bibinfo{author}{\bibfnamefont{R.}~\bibnamefont{Takahashi}},
  \bibinfo{author}{\bibfnamefont{J.}~\bibnamefont{Thornburg}},
  \bibnamefont{and}
  \bibinfo{author}{\bibfnamefont{J.}~\bibnamefont{Ventrella}},
  \bibinfo{journal}{Phys.\ Rev.\ Lett.} \textbf{\bibinfo{volume}{96}},
  \bibinfo{pages}{121101} (\bibinfo{year}{2006}).

\bibitem[{\citenamefont{Loffler et~al.}(2006)\citenamefont{Loffler, Rezzolla,
  and Ansorg}}]{Loffler2006}
\bibinfo{author}{\bibfnamefont{F.}~\bibnamefont{Loffler}},
  \bibinfo{author}{\bibfnamefont{L.}~\bibnamefont{Rezzolla}}, \bibnamefont{and}
  \bibinfo{author}{\bibfnamefont{M.}~\bibnamefont{Ansorg}},
  \bibinfo{journal}{Phys.\ Rev.\ D} \textbf{\bibinfo{volume}{74}},
  \bibinfo{pages}{104018} (\bibinfo{year}{2006}).

\bibitem[{\citenamefont{Herrmann et~al.}(2007)\citenamefont{Herrmann, Hinder,
  Shoemaker, and Laguna}}]{Herrmann2007b}
\bibinfo{author}{\bibfnamefont{F.}~\bibnamefont{Herrmann}},
  \bibinfo{author}{\bibfnamefont{I.}~\bibnamefont{Hinder}},
  \bibinfo{author}{\bibfnamefont{D.}~\bibnamefont{Shoemaker}},
  \bibnamefont{and} \bibinfo{author}{\bibfnamefont{P.}~\bibnamefont{Laguna}},
  \bibinfo{journal}{Class.\ Quantum Grav.} \textbf{\bibinfo{volume}{24}},
  \bibinfo{pages}{S33} (\bibinfo{year}{2007}).

\bibitem[{\citenamefont{Br{\"u}gmann et~al.}(2008)\citenamefont{Br{\"u}gmann,
  Gonz\'{a}lez, Hannam, Husa, Sperhake, and Tichy}}]{Bruegmann2006}
\bibinfo{author}{\bibfnamefont{B.}~\bibnamefont{Br{\"u}gmann}},
  \bibinfo{author}{\bibfnamefont{J.~A.} \bibnamefont{Gonz\'{a}lez}},
  \bibinfo{author}{\bibfnamefont{M.}~\bibnamefont{Hannam}},
  \bibinfo{author}{\bibfnamefont{S.}~\bibnamefont{Husa}},
  \bibinfo{author}{\bibfnamefont{U.}~\bibnamefont{Sperhake}}, \bibnamefont{and}
  \bibinfo{author}{\bibfnamefont{W.}~\bibnamefont{Tichy}},
  \bibinfo{journal}{Phys.\ Rev.\ D} \textbf{\bibinfo{volume}{77}},
  \bibinfo{eid}{024027} (\bibinfo{year}{2008}).

\bibitem[{\citenamefont{Scheel et~al.}(2008)\citenamefont{Scheel, Boyle, Chu,
  Kidder, Matthews, and Pfeiffer}}]{Scheel2008}
\bibinfo{author}{\bibfnamefont{M.~A.} \bibnamefont{Scheel}},
  \bibinfo{author}{\bibfnamefont{M.}~\bibnamefont{Boyle}},
  \bibinfo{author}{\bibfnamefont{T.}~\bibnamefont{Chu}},
  \bibinfo{author}{\bibfnamefont{L.~E.} \bibnamefont{Kidder}},
  \bibinfo{author}{\bibfnamefont{K.~D.} \bibnamefont{Matthews}},
  \bibnamefont{and} \bibinfo{author}{\bibfnamefont{H.~P.}
  \bibnamefont{Pfeiffer}} (\bibinfo{year}{2008}), \eprint{arXiv:0810.1767}.

\bibitem[{\citenamefont{Miller}(2005)}]{Miller2005}
\bibinfo{author}{\bibfnamefont{M.~A.} \bibnamefont{Miller}},
  \bibinfo{journal}{Phys. Rev. D} \textbf{\bibinfo{volume}{71}},
  \bibinfo{pages}{104016} (\bibinfo{year}{2005}).

\bibitem[{\citenamefont{Fairhurst}()}]{Fairhurst2008}
\bibinfo{author}{\bibfnamefont{S.}~\bibnamefont{Fairhurst}},
  \bibinfo{note}{{R}equired Waveform Accuracy, unpublished notes (2008)}.

\bibitem[{\citenamefont{Anderson et~al.}(2001)}]{T010095}
\bibinfo{author}{\bibfnamefont{S.}~\bibnamefont{Anderson}}
  \bibnamefont{et~al.}, \bibinfo{type}{Tech. Rep.}
  \bibinfo{number}{LIGO-T010095-00-Z}, \bibinfo{institution}{LIGO Project}
  (\bibinfo{year}{2001}),
  \urlprefix\url{http://www.ligo.caltech.edu/docs/T/T010095-00.pdf}.

\bibitem[{\citenamefont{Boyle et~al.}()\citenamefont{Boyle, Brown, Pekowsky,
  Pfeiffer, and Scheel}}]{Boyle2008b}
\bibinfo{author}{\bibfnamefont{M.}~\bibnamefont{Boyle}},
  \bibinfo{author}{\bibfnamefont{D.~A.} \bibnamefont{Brown}},
  \bibinfo{author}{\bibfnamefont{L.}~\bibnamefont{Pekowsky}},
  \bibinfo{author}{\bibfnamefont{H.~P.} \bibnamefont{Pfeiffer}},
  \bibnamefont{and} \bibinfo{author}{\bibfnamefont{M.~A.}
  \bibnamefont{Scheel}}, \bibinfo{note}{in preparation}.

\bibitem[{\citenamefont{Abbott et~al.}(2004{\natexlab{a}})}]{Abbott:2003yq}
\bibinfo{author}{\bibfnamefont{B.}~\bibnamefont{Abbott}} \bibnamefont{et~al.}
  (\bibinfo{collaboration}{LIGO Scientific}), \bibinfo{journal}{Phys. Rev. D}
  \textbf{\bibinfo{volume}{69}}, \bibinfo{pages}{082004}
  (\bibinfo{year}{2004}{\natexlab{a}}).

\bibitem[{\citenamefont{Jaranowski and Krolak}(1999)}]{Jaranowski:1998ge}
\bibinfo{author}{\bibfnamefont{P.}~\bibnamefont{Jaranowski}} \bibnamefont{and}
  \bibinfo{author}{\bibfnamefont{A.}~\bibnamefont{Krolak}},
  \bibinfo{journal}{Phys. Rev. D} \textbf{\bibinfo{volume}{59}},
  \bibinfo{pages}{063003} (\bibinfo{year}{1999}).

\bibitem[{\citenamefont{Dupuis and Woan}(2005)}]{Dupuis:2005xv}
\bibinfo{author}{\bibfnamefont{R.~J.} \bibnamefont{Dupuis}} \bibnamefont{and}
  \bibinfo{author}{\bibfnamefont{G.}~\bibnamefont{Woan}},
  \bibinfo{journal}{Phys. Rev. D} \textbf{\bibinfo{volume}{72}},
  \bibinfo{pages}{102002} (\bibinfo{year}{2005}).

\bibitem[{\citenamefont{Finn}(1992)}]{Finn1992}
\bibinfo{author}{\bibfnamefont{L.~S.} \bibnamefont{Finn}},
  \bibinfo{journal}{Phys.\ Rev.\ D} \textbf{\bibinfo{volume}{46}},
  \bibinfo{pages}{5236} (\bibinfo{year}{1992}).

\bibitem[{\citenamefont{Finn and Chernoff}(1993)}]{Finn1993}
\bibinfo{author}{\bibfnamefont{L.~S.} \bibnamefont{Finn}} \bibnamefont{and}
  \bibinfo{author}{\bibfnamefont{D.~F.} \bibnamefont{Chernoff}},
  \bibinfo{journal}{Phys.\ Rev.\ D} \textbf{\bibinfo{volume}{47}},
  \bibinfo{pages}{2198} (\bibinfo{year}{1993}).

\bibitem[{\citenamefont{Cutler and Flanagan}(1994)}]{CutlerFlanagan1994}
\bibinfo{author}{\bibfnamefont{C.}~\bibnamefont{Cutler}} \bibnamefont{and}
  \bibinfo{author}{\bibfnamefont{E.~E.} \bibnamefont{Flanagan}},
  \bibinfo{journal}{Phys.\ Rev.\ D} \textbf{\bibinfo{volume}{49}},
  \bibinfo{pages}{2658} (\bibinfo{year}{1994}).

\bibitem[{\citenamefont{Owen}(1996)}]{Owen:1995tm}
\bibinfo{author}{\bibfnamefont{B.~J.} \bibnamefont{Owen}},
  \bibinfo{journal}{Phys.\ Rev.\ D} \textbf{\bibinfo{volume}{53}},
  \bibinfo{pages}{6749} (\bibinfo{year}{1996}).

\bibitem[{\citenamefont{Abbott et~al.}(2004{\natexlab{b}})}]{Abbott:2003pj}
\bibinfo{author}{\bibfnamefont{B.}~\bibnamefont{Abbott}} \bibnamefont{et~al.}
  (\bibinfo{collaboration}{LIGO Scientific Collaboration}),
  \bibinfo{journal}{Phys. Rev. D} \textbf{\bibinfo{volume}{69}},
  \bibinfo{pages}{122001} (\bibinfo{year}{2004}{\natexlab{b}}).

\bibitem[{\citenamefont{Abbott et~al.}(2005)}]{Abbott:2005pe}
\bibinfo{author}{\bibfnamefont{B.}~\bibnamefont{Abbott}} \bibnamefont{et~al.}
  (\bibinfo{collaboration}{LIGO Scientific Collaboration}),
  \bibinfo{journal}{Phys. Rev. D} \textbf{\bibinfo{volume}{72}},
  \bibinfo{pages}{082001} (\bibinfo{year}{2005}).

\bibitem[{\citenamefont{Abbott et~al.}(2006)}]{Abbott:2005qm}
\bibinfo{author}{\bibfnamefont{B.}~\bibnamefont{Abbott}} \bibnamefont{et~al.}
  (\bibinfo{collaboration}{LIGO Scientific Collaboration and TAMA
  Collaboration}), \bibinfo{journal}{Phys.\ Rev.\ D}
  \textbf{\bibinfo{volume}{73}}, \bibinfo{pages}{102002}
  (\bibinfo{year}{2006}).

\bibitem[{\citenamefont{Abbott et~al.}(2008)}]{Abbott:2007xi}
\bibinfo{author}{\bibfnamefont{B.}~\bibnamefont{Abbott}} \bibnamefont{et~al.}
  (\bibinfo{collaboration}{{LIGO} Scientific Collaboration}),
  \bibinfo{journal}{Phys.\ Rev.\ D} \textbf{\bibinfo{volume}{77}},
  \bibinfo{pages}{062002} (\bibinfo{year}{2008}).

\bibitem[{\citenamefont{Apostolatos et~al.}(1994)\citenamefont{Apostolatos,
  Cutler, Sussman, and Thorne}}]{Apostolatos1994}
\bibinfo{author}{\bibfnamefont{T.~A.} \bibnamefont{Apostolatos}},
  \bibinfo{author}{\bibfnamefont{C.}~\bibnamefont{Cutler}},
  \bibinfo{author}{\bibfnamefont{G.~J.} \bibnamefont{Sussman}},
  \bibnamefont{and} \bibinfo{author}{\bibfnamefont{K.~S.}
  \bibnamefont{Thorne}}, \bibinfo{journal}{Phys.\ Rev.\ D}
  \textbf{\bibinfo{volume}{49}}, \bibinfo{pages}{6274 } (\bibinfo{year}{1994}).

\bibitem[{\citenamefont{Lazzarini and Weiss}(1995)}]{InitialLIGONoise}
\bibinfo{author}{\bibfnamefont{A.}~\bibnamefont{Lazzarini}} \bibnamefont{and}
  \bibinfo{author}{\bibfnamefont{R.}~\bibnamefont{Weiss}},
  \emph{\bibinfo{title}{{LIGO science requirements document}}}
  (\bibinfo{year}{1995}), \bibinfo{note}{{LIGO-E950018-02-E; See also}},
  \urlprefix\url{http://www.ligo.caltech.edu/~jzweizig/distribution/LSC_Data/}.

\bibitem[{Adv()}]{AdvancedLIGONoise}
\emph{\bibinfo{title}{{GWINC: Gravitational Wave Interferometer Noise
  Calculator}}}, \bibinfo{note}{v1 default parameters},
  \urlprefix\url{http://lhocds.ligo-wa.caltech.edu:8000/advligo/GWINC}.

\bibitem[{\citenamefont{Brown}(2008)}]{LIGO-G080178-04-Z}
\bibinfo{author}{\bibfnamefont{D.}~\bibnamefont{Brown}}, \bibinfo{journal}{for
  the LSC, LIGO document {LIGO}-G080178-04-Z}  (\bibinfo{year}{2008}),
  \urlprefix\url{http://www.ligo.caltech.edu/docs/G/G080178-04.pdf}.

\bibitem[{\citenamefont{Allen}(1996)}]{Allen1996}
\bibinfo{author}{\bibfnamefont{B.}~\bibnamefont{Allen}}, \bibinfo{journal}{LIGO
  Tech. Rep. {LIGO}-T960189-00-E}  (\bibinfo{year}{1996}),
  \urlprefix\url{http://www.ligo.caltech.edu/docs/T/T960189-00-E.pdf}.

\bibitem[{\citenamefont{Bose}()}]{Bose2005}
\bibinfo{author}{\bibfnamefont{S.}~\bibnamefont{Bose}},
  \emph{\bibinfo{title}{Effects of calibration inaccuracies: Applications in
  gravitational-wave detection and parameter estimation}},
  \bibinfo{note}{unpublished (2005)}.

\bibitem[{\citenamefont{Dietz et~al.}(2006)\citenamefont{Dietz, Garofoli,
  Gonzalez, Landry, O'Reilly, and Sung}}]{LIGOS4Calibration}
\bibinfo{author}{\bibfnamefont{A.}~\bibnamefont{Dietz}},
  \bibinfo{author}{\bibfnamefont{J.}~\bibnamefont{Garofoli}},
  \bibinfo{author}{\bibfnamefont{G.}~\bibnamefont{Gonzalez}},
  \bibinfo{author}{\bibfnamefont{M.}~\bibnamefont{Landry}},
  \bibinfo{author}{\bibfnamefont{B.}~\bibnamefont{O'Reilly}}, \bibnamefont{and}
  \bibinfo{author}{\bibfnamefont{M.}~\bibnamefont{Sung}}, \bibinfo{type}{Tech.
  Rep.} \bibinfo{number}{LIGO-T050262-01-D}, \bibinfo{institution}{{LIGO}
  Project} (\bibinfo{year}{2006}),
  \urlprefix\url{http://www.ligo.caltech.edu/docs/T/T050262-01.pdf}.

\bibitem[{\citenamefont{Barack and Cutler}(2004)}]{Barack2004}
\bibinfo{author}{\bibfnamefont{L.}~\bibnamefont{Barack}} \bibnamefont{and}
  \bibinfo{author}{\bibfnamefont{C.}~\bibnamefont{Cutler}},
  \bibinfo{journal}{\prd} \textbf{\bibinfo{volume}{70}},
  \bibinfo{pages}{122002} (\bibinfo{year}{2004}).

\bibitem[{\citenamefont{Sesana et~al.}(2005{\natexlab{a}})\citenamefont{Sesana,
  Haardt, Madau, and Volonteri}}]{Sesana:2005qn}
\bibinfo{author}{\bibfnamefont{A.}~\bibnamefont{Sesana}},
  \bibinfo{author}{\bibfnamefont{F.}~\bibnamefont{Haardt}},
  \bibinfo{author}{\bibfnamefont{P.}~\bibnamefont{Madau}}, \bibnamefont{and}
  \bibinfo{author}{\bibfnamefont{M.}~\bibnamefont{Volonteri}},
  \bibinfo{journal}{Class. Quant. Grav.} \textbf{\bibinfo{volume}{22}},
  \bibinfo{pages}{S363} (\bibinfo{year}{2005}{\natexlab{a}}).

\bibitem[{\citenamefont{Sesana et~al.}(2005{\natexlab{b}})\citenamefont{Sesana,
  Haardt, Madau, and Volonteri}}]{Sesana:2005}
\bibinfo{author}{\bibfnamefont{A.}~\bibnamefont{Sesana}},
  \bibinfo{author}{\bibfnamefont{F.}~\bibnamefont{Haardt}},
  \bibinfo{author}{\bibfnamefont{P.}~\bibnamefont{Madau}}, \bibnamefont{and}
  \bibinfo{author}{\bibfnamefont{M.}~\bibnamefont{Volonteri}},
  \bibinfo{journal}{Astrophys.\ J.} \textbf{\bibinfo{volume}{623}},
  \bibinfo{pages}{23} (\bibinfo{year}{2005}{\natexlab{b}}).

\bibitem[{\citenamefont{Buonanno et~al.}(2007)\citenamefont{Buonanno, Cook, and
  Pretorius}}]{Buonanno-Cook-Pretorius:2007}
\bibinfo{author}{\bibfnamefont{A.}~\bibnamefont{Buonanno}},
  \bibinfo{author}{\bibfnamefont{G.~B.} \bibnamefont{Cook}}, \bibnamefont{and}
  \bibinfo{author}{\bibfnamefont{F.}~\bibnamefont{Pretorius}},
  \bibinfo{journal}{Phys.\ Rev.\ D} \textbf{\bibinfo{volume}{75}},
  \bibinfo{eid}{124018} (\bibinfo{year}{2007}).

\end{thebibliography}
\end{document}